\documentclass[aps,prd,10pt,twocolumn,superscriptaddress,showpacs,longbibliography]{revtex4-1}

\pdfoutput=1
\usepackage{graphicx}
\usepackage{amsfonts,amsmath,amssymb,bm,bbm}
\usepackage{color}
\usepackage{enumitem}
\usepackage{url}
\usepackage{soul}

\usepackage{multirow}

\usepackage{hyperref}
\hypersetup{
    pdfnewwindow=true,     
    colorlinks=true,     
    linkcolor=blue,          
    citecolor=blue,      
    filecolor=blue,   
    urlcolor=blue    
}

\def\bi{\begin{itemize}[noitemsep,leftmargin=*]
\setlength\itemsep{1em}
        }
\def\ei{\end{itemize}}

\begin{document}

\title{Search for High-Energy Neutrino Emission from Radio-Bright AGN}

\author{Bei Zhou}
\email{beizhou@jhu.edu}
\thanks{\scriptsize \!\!  \href{https://orcid.org/0000-0003-1600-8835}{orcid.org/0000-0003-1600-8835}}
\affiliation{Department of Physics and Astronomy, Johns Hopkins University, Baltimore, Maryland 21218, USA}

\author{Marc Kamionkowski}
\email{kamion@jhu.edu}
\thanks{\scriptsize \!\!  \href{https://orcid.org/0000-0001-7018-2055}{orcid.org/0000-0001-7018-2055}}
\affiliation{Department of Physics and Astronomy, Johns Hopkins University, Baltimore, Maryland 21218, USA}

\author{Yun-feng Liang}
\email{liangyf@gxu.edu.cn}
\thanks{\scriptsize \!\!  \href{https://orcid.org/0000-0002-6316-1616}{orcid.org/0000-0002-6316-1616}}
\affiliation{Laboratory for Relativistic Astrophysics, Department of Physics, Guangxi University, Nanning 530004, China}

\date{June 16, 2021}

\begin{abstract}
We investigate the possibility that radio-bright active galactic nuclei (AGN) are responsible for the TeV--PeV neutrinos detected by IceCube. We use an unbinned maximum-likelihood-ratio method, 10 years of IceCube muon-track data, and 3388 radio-bright AGN selected from the Radio Fundamental Catalog. None of the AGN in the catalog have a large global significance. The two most significant sources have global significance of $\simeq$ 1.5$\sigma$ and 0.8$\sigma$, though 4.1$\sigma$ and 3.8$\sigma$ local significance. Our stacking analyses show no significant correlation between the whole catalog and IceCube neutrinos.  We infer from the null search that this catalog can account for at most 30\% (95\% CL) of the diffuse astrophysical neutrino flux measured by IceCube. Moreover, our results disagree with recent work that claimed a 4.1$\sigma$ detection of neutrinos from the sources in this catalog, and we discuss the reasons of the difference.

\end{abstract}

\maketitle


\section{Introduction}
\label{sec:introduction}

The TeV--PeV diffuse astrophysical neutrinos detected by IceCube~\cite{Aartsen:2013jdh, Aartsen:2015knd, Stettner:2019tok, Stachurska:2019wfb, Aartsen:2020aqd, Abbasi:2020jmh} have opened a new window for astrophysics~\cite{Kistler:2006hp, Beacom:2007yu, Murase:2010cu, Murase:2013ffa, Murase:2013rfa, Ahlers:2014ioa, Tamborra:2014xia, Murase:2014foa, Bechtol:2015uqb, Kistler:2016ask, Bartos:2016wud, Sudoh:2018ana, Bartos:2018jco, Bustamante:2019sdb, Hovatta:2020lor} and provide a new high-energy-particle source for elementary particle physics~\cite{Hooper:2002yq, Borriello:2007cs, Connolly:2011vc, Klein:2013xoa, Gauld:2015kvh, Aartsen:2017kpd, Bustamante:2017xuy, Cornet:2001gy, Lipari:2001ds, AlvarezMuniz:2002ga, Beacom:2002vi, Han:2003ru, Han:2004kq, Hooper:2004xr, Hooper:2005jp, GonzalezGarcia:2005xw, Ng:2014pca, Ioka:2014kca, Bustamante:2016ciw, Salvado:2016uqu, Coloma:2017ppo, Zhou:2019vxt, Beacom:2019pzs, IceCube:2021rpz}.  However, the origin of these neutrinos remains a mystery, albeit one that is thought to be related to the origin of high- and ultrahigh-energy cosmic rays~\cite{Halzen:2008zj, Aartsen:2013dsm, Kistler:2013my, Anchordoqui:2013qsi, Fang:2013vla, Aartsen:2016ngq}, which still remains obscure. In the past decade, significant efforts have been made to search for the sources of these astrophysical neutrinos~\cite{Abbasi:2010rd, Aartsen:2013uuv, Aartsen:2014cva, Aartsen:2015wto, Aartsen:2016oji, Aartsen:2018ywr, Aartsen:2019fau, Abbasi:2020dfi}. A few sources were found to be very likely TeV--PeV neutrino emitters, including TXS 0506+056~\cite{IceCube:2018cha, IceCube:2018dnn} and NGC 1068~\cite{Aartsen:2019fau}. The blazar TXS 0506+056 was found to have a 3$\sigma$ (global significance) spatial and temporal association with a high-energy muon-track event induced by a $\sim 300$~TeV neutrino~\cite{IceCube:2018dnn}. This led to the discovery of a neutrino flare between 2014 and 2015 from the location of TXS 0506+056, showing 3.5$\sigma$ global significance~\cite{IceCube:2018cha}. Note that these two measurements are statistically independent. The Seyfert II galaxy NGC 1068 showed an excess of 2.9$\sigma$ (global significance) in the time-integrated source searches using 10 years of IceCube data~\cite{Aartsen:2019fau}.

Still, the majority of the diffuse astrophysical neutrinos remain largely unexplained. Many different types of sources have been proposed and searched with data, including gamma-ray bursts~\cite{Waxman:1997ti, Abbasi:2009ig, Abbasi:2011qc, Abbasi:2012zw, Aartsen:2014aqy, Aartsen:2016qcr, Aartsen:2017wea}, gamma-ray blazars~\cite{Aartsen:2016lir, Hooper:2018wyk, Yuan:2019ucv, Luo:2020dxa, Smith:2020oac}, choked jet supernovae~\cite{Senno:2017vtd, Esmaili:2018wnv}, pulsar wind nebulae~\cite{Aartsen:2020eof}, etc. However, none of these searches found a strong correlation. For example, motivated by theoretical considerations~\cite{Atoyan:2001ey, Essey:2010er, Protheroe:1996uu} and TXS 0506+056~\cite{IceCube:2018cha, IceCube:2018dnn}, searches for high-energy neutrino emission from gamma-ray blazars were implemented~\cite{Aartsen:2016lir, Hooper:2018wyk, Smith:2020oac}. However, they were found to contribute at most $\sim 20\%$ to the diffuse neutrino flux. (See also Ref.~\cite{Yuan:2019ucv} for a general discussion.)
On the other hand, it has been shown that there exist high-energy neutrino sources hidden in GeV--TeV gamma rays~\cite{Murase:2015xka, Capanema:2020rjj, Capanema:2020oet}. Otherwise, the gamma rays from all the transparent sources would overshoot the isotropic diffuse gamma-ray flux measured by {\it Fermi}-LAT ~\cite{Ackermann:2014usa}. Therefore, other categories of sources must be investigated, including those not characterized by gamma rays.

Recently, the possibility that radio-bright AGN with strong parsec-scale cores are associated with IceCube neutrino events has been considered~\cite{Plavin:2020emb, Plavin:2020mkf}. The 3388 radio-bright AGN that were used for the analysis form a complete catalog and were selected from the Radio Fundamental Catalog~\cite{RFCweb}, which collects VLBI (very-long-baseline interferometry) observations since 1980. These radio-bright AGN have very diverse gamma-ray properties. In Ref.~\cite{Plavin:2020emb}, a $3.1\sigma$ significance was found when 56 muon-track events ($>200$~TeV) from 2009 to 2019 were considered. In Ref.~\cite{Plavin:2020mkf}, a pretrial p-value sky map published by the IceCube Collaboration~\cite{IC_pvalue_web} was found to have a $3\sigma$ correlation with radio-bright AGN. The sky map was obtained from the analysis of muon-track data between 2008 and 2015. The combination of the two analyses yields a $4.1\sigma$-significance detection~\cite{Plavin:2020mkf}. Moreover, the sources with stronger X-band ($\simeq 8$ GHz) flux densities show more significant correlations, implying that the X-band flux could be an indicator of TeV--PeV neutrino emission. 

\begin{figure*}[t!]
\includegraphics[width=0.99\textwidth]{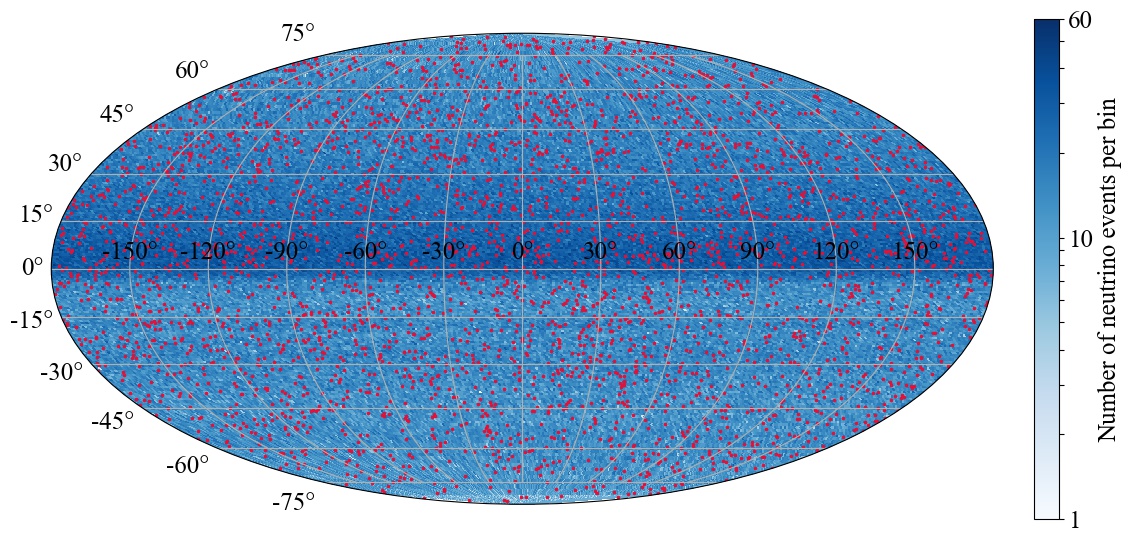}
\caption{
Distributions of the 3388 radio-bright AGN ({\it red}) and 1134450 neutrino events ({\it blue}) used in this work on the celestial sphere.  The sky map is in the equatorial coordinate system, with the {\it horizontal} and {\it vertical} axes representing right ascension (RA) and declination (Dec), respectively. To IceCube, the upper half ($\text{Dec} > 0^\circ$) is the northern sky (seen through Earth), while the lower half ($\text{Dec} < 0^\circ$) is the southern sky. The neutrino events are shown as numbers in $360 \times 180$ of $\text{RA} \times \text{sin}\text{(Dec)}$ bins, so that all the bins have the same solid angle. See Sec.~\ref{sec_data} for details.
}
\label{fig_skymap}
\end{figure*}

In this paper, we revisit the correlation between these radio-bright AGN and the TeV--PeV astrophysical neutrinos detected by IceCube. We use 10 years of IceCube muon-track data (April 2008 to July 2018)~\cite{Abbasi:2021bvk, data_webpage}, made publicly available just recently, and an unbinned maximum-likelihood-ratio method~\cite{Braun:2008bg, Braun:2009wp, Abbasi:2010rd, Aartsen:2013uuv}.  The method, which has been extensively used for analyzing IceCube data~\cite{Abbasi:2010rd, Aartsen:2013uuv, Aartsen:2016lir, Hooper:2018wyk, Aartsen:2020eof, Smith:2020oac}, takes into account the information provided by every single event. It is also widely used in other types of astroparticle measurements (e.g., Refs.~\cite{Bays:2011si, Ackermann:2015lka, Ahnen:2016qkx, Abdallah:2018qtu}).  The conclusion of our analysis is that there is no significant correlation between IceCube neutrinos and these 3388 radio-bright AGN.  Our null search suggests that no more than $30\%$ of the IceCube neutrino flux may have come from these sources.

This paper is organized as follows. In Sec.~\ref{sec_data}, we describe the radio-bright AGN (Sec.~\ref{sec_data_srcs}) and the 10 years of IceCube muon-track data (Sec.~\ref{sec_data_nus}). Sec.~\ref{sec_llh} discusses the unbinned maximum-likelihood-ratio method used for our analysis. Sec.~\ref{sec_results} shows the results for both the analysis that searches for neutrino emission from every source location in the catalog (Sec.~\ref{sec_SglSrcAnlys}) and the analysis stacking all the sources together for the correlation with all-sky neutrinos (Sec.~\ref{sec_StackingAnlys}). We also compare our results with previous work (Sec.~\ref{sec_cmpr2prevWk}). We conclude in Sec.~\ref{sec_conclusions}.


\section{Data}
\label{sec_data}

\subsection{Radio-bright AGN catalog from VLBI observations}
\label{sec_data_srcs}

The radio-bright AGN used in this work are selected from the Radio Fundamental Catalog~\cite{RFCweb} in the Astrogeo database~\cite{astrogeoWeb}. The Radio Fundamental Catalog collects archival VLBI flux densities at several frequencies between 2 and 22 GHz as well as precise positions with milliarcsecond accuracies for more than $10^5$ compact radio sources. It uses all available VLBI (very-long-baseline interferometry) observations under absolute astrometry and geodesy programs since April 11, 1980. The catalog is regularly updated with more sources and more observations per source. In this work, we use the version ``rfc\_2019c'', which comprises 16466 objects. 

Moreover, as in Ref.~\cite{Plavin:2020emb}, we obtain a complete catalog of radio-bright AGN by selecting the sources in the Radio Fundamental Catalog {\it with X-band ($\simeq 8$ GHz) flux densities larger than 0.15~Jy}. This gives us 3388 sources, with the total X-band flux density $\simeq 1518$~Jy. This is the only deep statistically complete catalog that can be obtained from the Radio Fundamental Catalog (see Ref.~\cite{Plavin:2020emb} for details).  Moreover, these radio-bright AGN have very diverse gamma-ray properties. Fig.~\ref{fig_skymap} shows the distribution of the 3388 sources on the celestial sphere. Their locations are nearly isotropic.

\subsection{Neutrino data from IceCube}
\label{sec_data_nus}
The IceCube Neutrino Observatory detects neutrinos by detecting the Cherenkov photons emitted by relativistic secondary charged particles from neutrino interactions in and outside the detector~\cite{Achterberg:2006md, ICweb}. Two basic kinds of event topologies are formed in the detector. One is an elongated track formed by muons due to their low energy-loss rate in matter. The other is a shower which looks like a round and big blob formed by electrons (electromagnetic shower) or hadrons (hadronic shower). The track events have a better angular resolution (as good as $< 1^\circ$) while worse energy resolution ($\sim 200\%$ at $\sim 100$~TeV) than the shower events ($\sim$ $10^\circ$--$15^\circ$ and $\sim 15\%$ above 100 TeV)~\cite{ICres}. Therefore, the track events are suited to search for point sources.

There are two kinds of track events, through-going and starting tracks. The through-going tracks are those with parent neutrinos that have interacted outside the detector, while inside for starting tracks. Overall, the through-going tracks have better angular resolution and $\sim 10$ times larger effective area than that of starting tracks~\cite{Aartsen:2016oji}.

The data released by the IceCube Collaboration span from April 2008 to July 2018~\cite{Abbasi:2021bvk, data_webpage}. The same data have been used in IceCube's 10-year time-integrated neutrino point-source search~\cite{Aartsen:2019fau}. In total, there are 1134450 muon-track events. The information for each track is provided. These ten years of data can be grouped into five samples corresponding to different construction phases of IceCube, including 1) IC40, 2) IC59, 3) IC79, 4) IC86-I, and 5) IC86-II to IC86-VII. The numbers in the names represent the numbers of strings in the detector on which digital optical modules are deployed.

There are two main features in the new dataset that are different from the previously released three years of data from June 2010 to April 2012~\cite{Abbasi:2021bvk, data_webpage}. First, the reconstructed directional uncertainty assumes a lower limit of $0.2^\circ$ in order to avoid unaccounted-for uncertainties and to ensure no single event dominates the likelihood analysis. Second is the given uptime, which contains lists of time periods of good runs for each season.

Fig.~\ref{fig_skymap} shows the distribution of the muon-track events on the celestial sphere in the equatorial coordinate system. The distribution is shown as numbers in $360 \times 180$ of $\text{RA} \times \text{sin(Dec)}$ bins, so that all the bins have the same solid angle. More events came from the northern sky (upper half; $\text{Dec} > 0^\circ$) than the southern sky (lower half; $\text{Dec} < 0^\circ$), because the latter has a higher energy threshold to the events to reduce the contamination from atmospheric muons~\cite{Abbasi:2021bvk}. For the southern sky, the events are nearly evenly distributed. For the northern sky, there are more events as it is closer to $\text{Dec} = 0^\circ$ because of the absorption of neutrinos by Earth.  For more details, see Refs.~\cite{Abbasi:2021bvk, data_webpage}.


\section{Analysis formalism}
\label{sec_llh}

Our analysis uses an unbinned maximum-likelihood-ratio method~\cite{Braun:2008bg, Braun:2009wp, Abbasi:2010rd, Aartsen:2013uuv}. The likelihood function is given by the product of probability density functions (PDFs) of each muon-track event (indexed by $i$) in the five data samples (indexed by $k$):
\begin{equation}
\mathcal{L} \left(n_s \right) = \prod_{k} \prod_{i \in k}  \left[ \frac{n_s^k}{N_k} S_{i}^k + \left(1-\frac{n_s^k}{N_k}\right) B_i^k \right] \,,
\label{eq_llh_func}
\end{equation}
where $n_s$ is the total number of events from the source(s) combining all five samples, which is to be fit in our analysis, and $N_k$ is the total number of events in sample $k$. The $S_i^k$ and $B_i^k$ are the signal and background PDFs of event $i$ in the sample $k$, representing the probability of the event coming from the source(s) or background, respectively. Note the $k$ in the superscript is not an exponent but an index. Moreover, $n_s^k$ is the number of expected signal events coming from sample $k$, which is determined by
\begin{equation}
n_s^k = f_k \times n_s \,,
\label{eq_nsk}
\end{equation}
where $f_k$ is the expected fractional contribution from sample $k$, given by Eqs.~(\ref{eq_fk_1}) and (\ref{eq_fk_2}).

We calculate the background PDFs directly from data. For each sample, given the Dec $\delta_i$ of an event, the background PDF is determined by the relative number of events in $\delta_i \pm 3^\circ$ divided by the solid angle. We do not use events with $|\delta_i| > 87^\circ$ (only account for $\simeq 0.1\%$ of the sky and the total events). The calculated background PDFs have precision of less than 2\% for most directions. The dependence on RA is negligible because IceCube, located at the South Pole, has a nearly uniform acceptance in RA.

\begin{figure}[t!]
\includegraphics[width=\columnwidth]{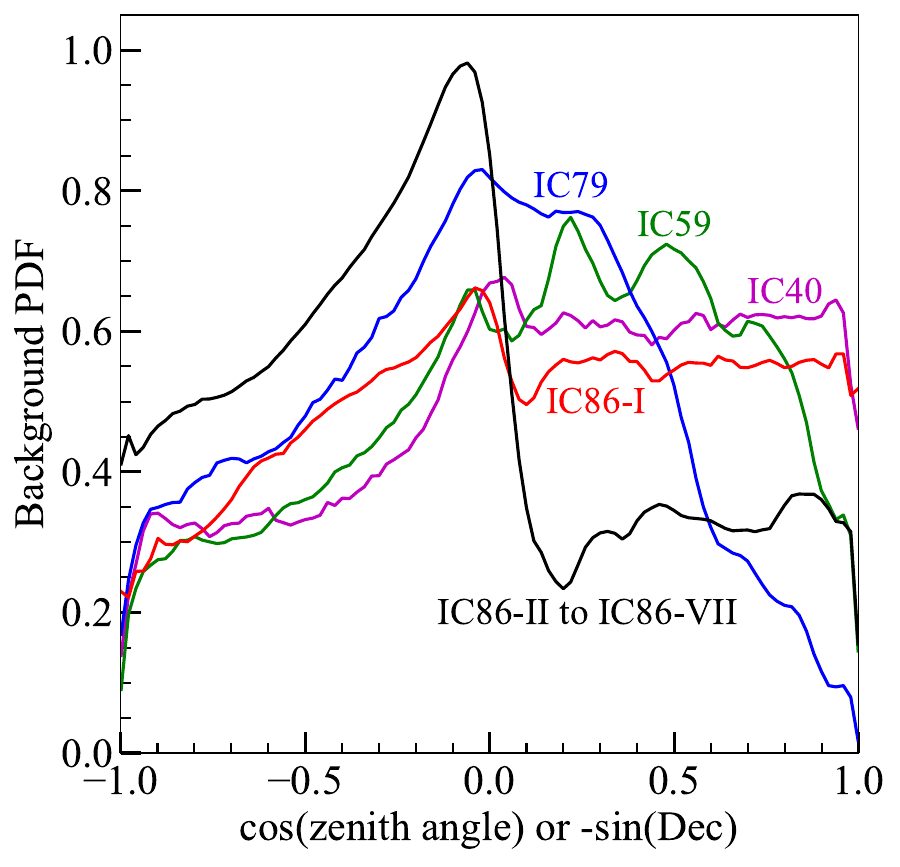}
\caption{RA-integrated background PDF for different phases of IceCube, as labeled. Shown are IC40 ($\simeq$ 1-year exposure, 36900 events), IC59 ($\simeq$ 1-year exposure, 107011 events), IC79 ($\simeq$ 1-year exposure, 93133 events), IC86-I ($\simeq$ 1-year exposure, 136244 events), and IC86-II to IC86-VII ($\simeq$ 6-year exposure, 761162 events). The left half is the northern sky, while right half the southern sky with respect to IceCube.
}
\label{fig_bkgdpdf}
\end{figure}

Fig.~\ref{fig_bkgdpdf} shows the RA-integrated background PDFs. They are consistent with the distribution of muon-track events in Fig.~\ref{fig_skymap} (see also Sec.~\ref{sec_data_nus}).

The signal PDF is expressed as a 2D symmetric Gaussian spatial distribution, 
\begin{equation}
S_{ij}^k \equiv S_k \left( \vec{x}_i, \sigma_i, \vec{x}_j \right) 
= 
\frac{1}{2 \pi \sigma_i^{2}} \exp\left( -\frac{ D( \vec{x}_i, \vec{x}_j )^2 } {2 \sigma_i^{2} }\right) 
\label{eq_Gaussian}
\end{equation}
where $\vec{x}_i$ and $\vec{x}_j$ are the coordinates of the event $i$ and source $j$ on the celestial sphere, respectively, and $D( \vec{x}_i, \vec{x}_j )$ is the angular distance between them.  Here, $\sigma_i$ is the uncertainty of the reconstructed direction of the track (much larger than the scattering angle between the incoming neutrino and the outgoing muon~\cite{Aartsen:2014cva}).  We do not include the energy part of the signal PDF because of lower statistics in the published smearing matrices~\cite{Abbasi:2021bvk, data_webpage}. However, our analysis is sensitive to different neutrino spectrum through the energy- and zenith-dependent effective area~\cite{Abbasi:2021bvk, data_webpage}.

For the case of multiple sources, the signal PDF is a weighted sum of the signal PDFs ($S^k_{ij}$) of all the sources, indexed by $j$~\cite{Abbasi:2010rd, Aartsen:2013uuv},
\begin{equation}
S_i^k
= 
\frac{\sum_{j} w_{j, \rm model} w^k_{j, \rm acc} S_{ij}^k}{\sum_{j} w_{j,\rm model} w^k_{j, \rm acc}} ,
\label{eq_sgnpdf_stacking}
\end{equation} 
where the $w_j$ are weighting factors which are proportional to the number of expected events from source $j$. The $w^k_{j, \text{acc}}$ reflects the detector's acceptance, which is proportional to the convolution of detector effective area ($A^k_\text{eff}$) and neutrino spectrum, 
\begin{equation}
w^k_{j, \rm acc} (\delta_j)
\propto t_k \times \int A^k_\text{eff}(E_\nu, \delta_j) E_\nu^{\Gamma} \, d E_\nu \,,
\end{equation}
where $\delta_j$ is the Dec of the source, $t_k$ the total detector uptime of the sample $k$, $E_\nu$ neutrino energy, and $\Gamma$ is the spectral index of the signal neutrinos, which is assumed to be the same for all the sources.

The $w_{j, \rm model}$ describes the expected relative number of events due to the intrinsic properties of the sources, as discussed in Sec.~\ref{sec_wjm}. 

The $f_k$ in Eq.~(\ref{eq_nsk}) can therefore be calculated by 
\begin{equation}
f_k = 
\frac{ w^k_{j, \rm acc} }{\sum_{k} w^k_{j, \rm acc}} ,
\label{eq_fk_1}
\end{equation}
for one source, or 
\begin{equation}
f_k = 
\frac{ \sum_{j} w_{j, \rm model} w^k_{j, \rm acc} }{\sum_{k} \sum_{j} w_{j,\rm model} w^k_{j, \rm acc}} ,
\label{eq_fk_2}
\end{equation}
for multiple sources.

The test statistic (TS) of our analysis is,
\begin{equation}
\text{TS}(n_s) = 2 \ln{ \frac{ \mathcal{L}(n_s) }{ \mathcal{L}(n_s = 0) } }\,,
\label{eq_TS}
\end{equation}
where the denominator is the background or null hypothesis that all the events come from background. The best-fit number $\hat{n}_s$ of signal events is obtained by maximizing the TS value ($\rm TS_{max}$). 

If the background hypothesis is true, the probability distribution for ${\rm TS_{max}}$ is approximately a $\chi^2$ distribution, i.e.,
\begin{equation}
\text{PDF($\rm TS_{max}$)} \simeq \chi^2_1 ({\rm TS_{max}}) \,,
\label{eq_wilks}
\end{equation}
when the number of observations ($N_k$ for our case) is large (Wilks' theorem~\cite{Wilks1938}). The subscript ``1'' is the degree of freedom of the $\chi^2$ distribution, determined by the difference in the number of free parameters in the signal hypothesis and background hypothesis. Therefore, the $\sqrt{ {\rm TS_{max}} }$ is a good approximation of the significance for rejecting the background hypothesis, as it follows the standard normal distribution. We also verify in Sec.~\ref{sec_results} the use of Wilks' theorem in our analysis by comparing with simulations.

\begin{table*}[ht!]
\caption{ \label{tab_list_srcs} List of the five sources with highest significance. 
}
\medskip
\renewcommand{\arraystretch}{1.1} 
\centering
\begin{tabular*}{0.98\textwidth}{c|c|c|c|c|c|c}
IVS name	&	J2000 name	&	X-band flux density (Jy)	&	 $\hat{n}_s$	&	TS$_{\rm max}$	&	Pretrial p value, significance	&	Post-trial p value, significance	\\	\hline \hline
1303-170	&	J1306-1718	&	$ 0.208  $	&	21.6	&	16.6	&	$2.28\times10^{-5}$, 4.1$\sigma$	&	0.074, 1.5$\sigma$	\\	\hline
2245+029	&	J2247+0310	&	$ 0.434 $	&	50.8	&	14.5	&	$7.14\times10^{-5}$, 3.8$\sigma$	&	0.21, 0.8$\sigma$ 	\\	\hline
0228-163	&	J0231-1606	&	$ 0.162 $	&	15.9	&	9.8	&	$8.90\times10^{-4}$, 3.1$\sigma$	&	0.95, 0	\\	\hline
1424+240	&	J1427+2348	&	$ 0.187 $	&	38.1	&	8.9	&	$1.42\times10^{-3}$, 3.0$\sigma$	&	0.99, 0	\\	\hline
0958+559	&	J1001+5540	&	$ 0.180 $	&	27.2	&	8.3	&	$2.02\times10^{-3}$, 2.9$\sigma$	&	1.0, 0	\\	\hline
\end{tabular*}
\end{table*}

\section{Results and Discussions}
\label{sec_results}

We now present our analysis with the catalog of radio-bright AGN, using the formalism in Sec.~\ref{sec_llh} and 10 years of IceCube data. We begin in Sec.~\ref{sec_SglSrcAnlys} by analyzing the significance of each source location in the catalog. Then, in Sec.~\ref{sec_StackingAnlys}, we study the correlation of all-sky diffuse astrophysical neutrinos with the whole catalog by stacking the sources together in the likelihood analysis [Eq.~(\ref{eq_sgnpdf_stacking})]. We have checked our analysis code with relevant analyses in previous work~\cite{Aartsen:2016lir, IceCube:2018cha, IceCube:2018dnn, Hooper:2018wyk, Aartsen:2019fau, Smith:2020oac}, and we get consistent results. Finally, in Sec.~\ref{sec_cmpr2prevWk}, we compare our result with previous work~\cite{Plavin:2020emb, Plavin:2020mkf}.

\subsection{Single-source analysis} 
\label{sec_SglSrcAnlys}

\begin{figure}[t!]
\includegraphics[width=\columnwidth]{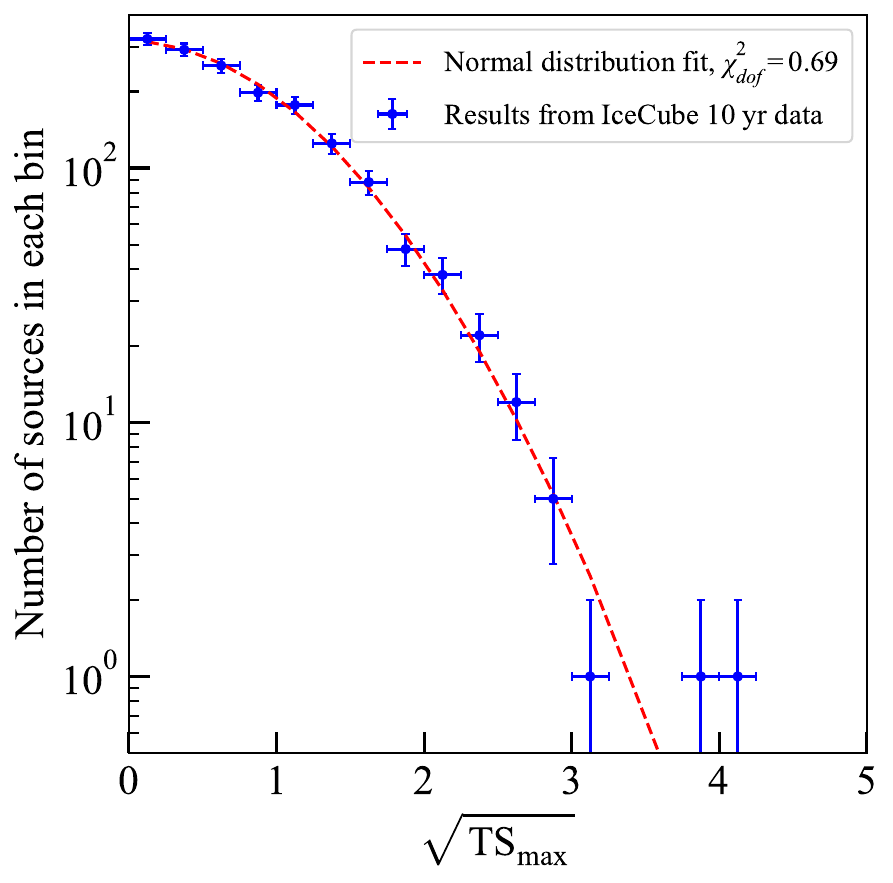}
\caption{
Distribution of maximized TS values for the radio-bright AGN with $\hat{n}_s \ge 0$ from our likelihood analysis with 10 years of IceCube muon-track data. The error bar is the square root of the number in each bin.  The red dashed line is the best normal distribution fit, which is proportional to the standard normal distribution.
}
\label{fig_TS_dist}
\end{figure}

Fig.~\ref{fig_TS_dist} shows the $\sqrt{\rm TS_{max}}$ for radio-bright AGN in the catalog. We plot it as the number of sources in each $\sqrt{\rm TS_{max}}$ bin, with the error bar given by the square root of the number. We use a normal distribution to fit the numbers with error bars. The best fit is consistent with the standard normal distribution times the number of sources divided by the number of bins per unit x axis and has a reduced chi square found to be 0.69. Therefore, the probability density distribution of $\sqrt{\rm TS_{max}}$ of the sources in this catalog follows the standard normal distribution. According to Wilks' theorem~\cite{Wilks1938} (Sec.~\ref{sec_llh}), this means that at the positions of most sources the background hypotheses are favored, implying no very strong correlation between diffuse astrophysical neutrinos and this catalog. This is further and quantitatively supported by our stacking analysis in Sec.~\ref{sec_StackingAnlys}.

Table~\ref{tab_list_srcs} lists the five sources with the highest $\text{TS}_\text{max}$, from which we calculate the pretrial (local) p value, $p_\text{lcoal}$, and the corresponding significance in units of standard normal deviations. The post-trial (global) p value can be calculated by $p_\text{global} = 1 - (1 - p_\text{local})^{N_\text{src}}  $, from which we get the post-trial significance. Here the $N_\text{src}$ is the number of sources. The source with the highest post-trial significance is J1306-1718, which has $\simeq 1.5\sigma$ ($p_\text{global} = 0.074$).  All the other sources have post-trial significance less than 1. The X-band flux densities of the sources with highest local significance are not high compared to other sources in the catalog, implying that the X-band flux densities of radio-bright AGN might not be an indicator of high-energy emission. Moreover, we do not find anything special about these sources with high local significance compared to other sources in the catalog.

To check the pretrial p value (and significance) calculated from Wilks' theorem~\cite{Wilks1938}, we simulate $N_\text{src}^\text{simu}$ ($> 10^6$) random source locations on the celestial sphere following uniform distributions in RA and $\sin$(Dec) [= $-\cos$(zenith angle) for IceCube location]. For each location, we maximize the TS value. The p value of a real source in the catalog is then determined by the number of the simulated locations with $\text{TS}_\text{max}$ larger than that of the real source ($M_\text{src}^\text{simu}$) divided by $N_\text{src}^\text{simu}$, i.e., $M_\text{src}^\text{simu}$/$N_\text{src}^\text{simu}$. We find that the p values calculated from Wilks' theorem and simulation are consistent with each other within uncertainties ($\simeq 1/\sqrt{M_\text{src}^\text{simu}}$).

\begin{figure*}[t!]
\includegraphics[width=0.465\textwidth]{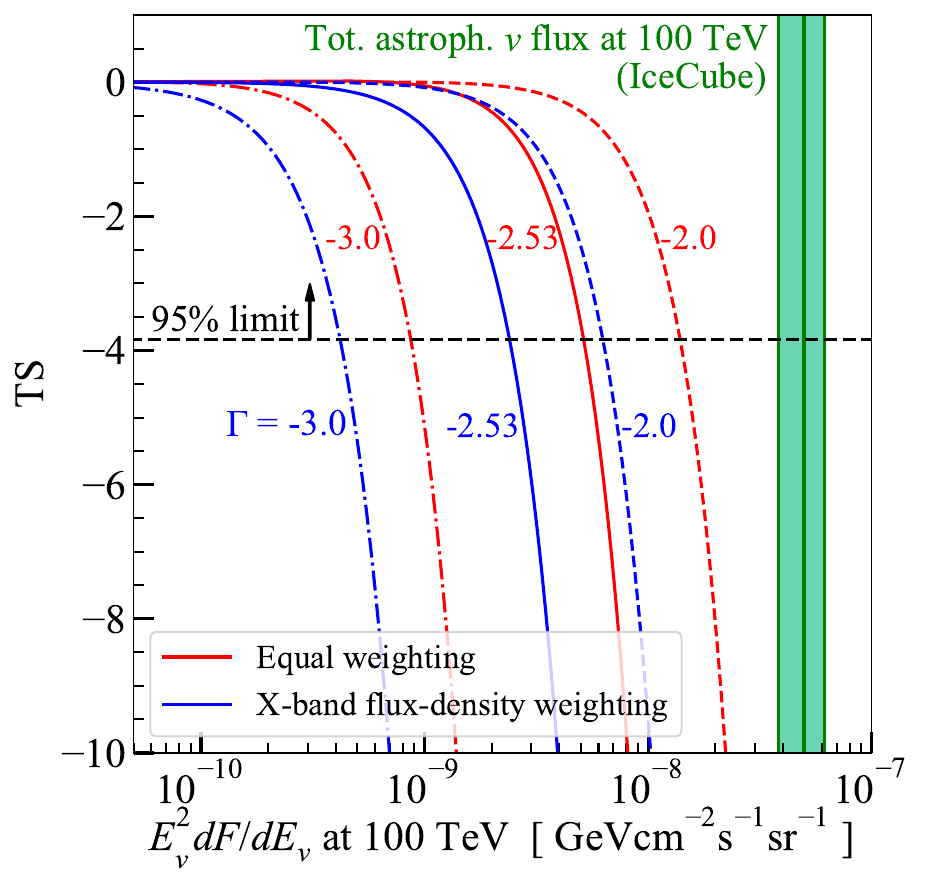}
\includegraphics[width=0.49\textwidth]{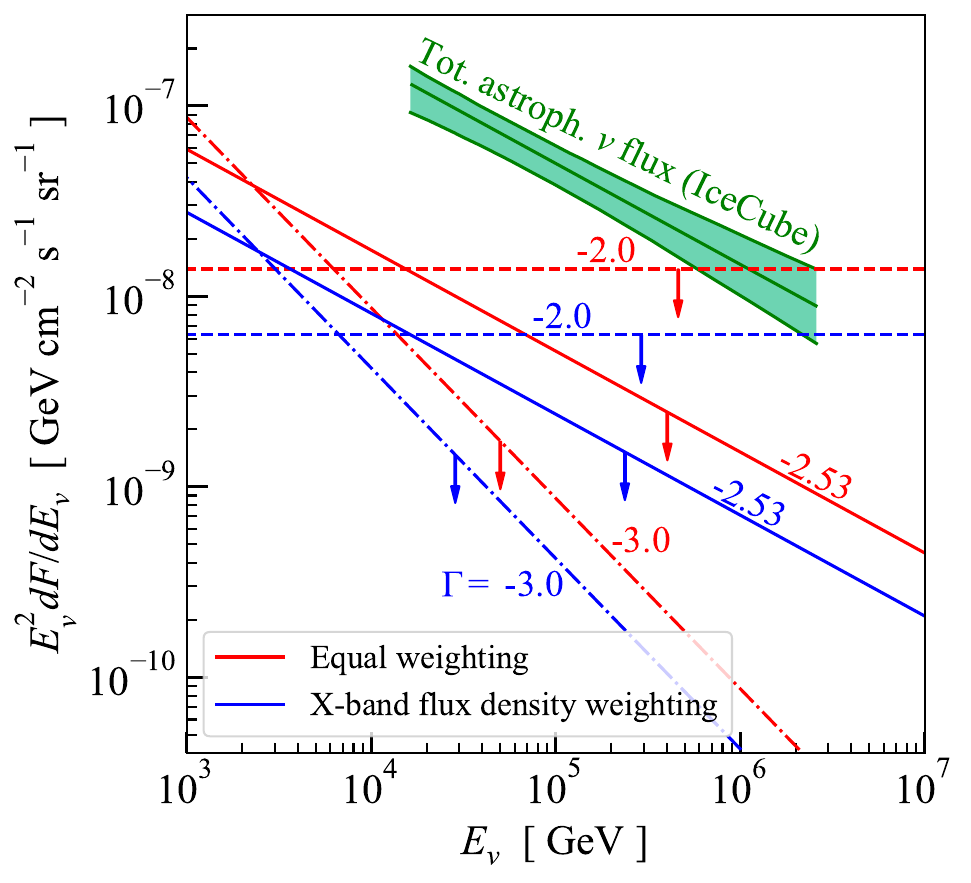}
\caption{
Our results of all six-flavor neutrino fluxes combining all the sources in the catalog for the two weighting schemes (red and blue lines), each with three spectra indices as labeled in each panel. Also shown is all six-flavor diffuse astrophysical neutrino flux measured by IceCube~\cite{Aartsen:2015knd, Aartsen:2020aqd} with uncertainties (green band; the three green lines represent lower limits, center value, and upper limits). Both our results and IceCube measurements assume equal flux for all six flavors. Both panels use the same line styles. 
{\it Left:} TS values as a function of the energy fluxes at 100~TeV.  The black dashed horizontal line shows where we set the upper limits at 95\% confidence level on each case, corresponding to $\text{TS} \simeq -3.84$. 
{\it Right:}  the energy fluxes versus neutrino energy. Note in the right panel the red and blue lines are {\it upper limits} at 95\% confidence level.
}
\label{fig_stacking} 
\end{figure*}

\subsection{Stacking analysis}
\label{sec_StackingAnlys}

The previous subsection shows that none of the sources in the catalog show significant neutrino emission, and the distribution of their $\text{TS}_\text{max}$ strongly favors the background hypothesis. In this subsection, we do a stacking analysis which integrates possible signals from all the sources. Then, we derive their muon neutrino ($\nu_\mu + \bar{\nu}_\mu$) flux by 
\begin{equation}
\begin{aligned}
\hat{n}_s =  2\pi \sum_{k} \, t_k  \int  d  \sin & \, \delta  \int A^k_\text{eff}(E_\nu, \delta) \,  \frac{d F}{d E_\nu}(E_\nu) \, d E_\nu \,,
\end{aligned}
\end{equation}
where $ d F(E_\nu) / d E_\nu =  \Phi_0 \times \left({E_\nu}/{100\ \text{TeV}}\right)^{\Gamma}$ and $\Phi_0$ is the flux normalization at 100 TeV. We also compare their flux with the diffuse astrophysical neutrino flux measured by IceCube~\cite{Aartsen:2015knd, Aartsen:2020aqd}.

\subsubsection{Weighting schemes, $w_{j,\rm model}$ }
\label{sec_wjm}

We use two weighting schemes [Eq.~(\ref{eq_sgnpdf_stacking})]. For the first, we assume that the high-energy neutrino fluxes of the sources are independent of their intrinsic properties. This leads to
\begin{equation}
w_{j,\rm model} = 1 \, .
\end{equation}

For the second scheme, we assume that the high-energy neutrino fluxes are proportional to the X-band (radio) flux densities, i.e., 
\begin{equation}
w_{j,\rm model} = \text{[X-band flux density]} \, .
\end{equation}
This is motivated by the findings in Refs.~\cite{Plavin:2020emb, Plavin:2020mkf} that sources with higher X-band flux densities contribute more to the correlation significance and that the significance increases when a subsample with a higher threshold of X-band flux density is used. Therefore, in this scheme, sources brighter in the X band have higher weights in their contribution to the signal PDF [Eq.~(\ref{eq_sgnpdf_stacking})].

\subsubsection{Results of the stacking analysis}

Fig.~\ref{fig_stacking} shows the results of our stacking analysis. {\it For both panels}, our results of all six-flavor neutrino flux combining all the sources in the catalog are shown in red (for equal weighting) and blue lines (for X-band flux-density weighting), each with three different spectral indices. The choice of the spectral indices is explained below. Also shown for comparison is the all six-flavor astrophysical neutrino flux measured by IceCube with uncertainty~\cite{Aartsen:2015knd, Aartsen:2020aqd}. Both our results and IceCube measurements assume equal flux for all six flavors. 

{\it The left panel} shows the TS value as a function of the energy fluxes at 100~TeV, which is about the median energy in the IceCube measurement range (see the right panel). If other energies are chosen, the curves shift only horizontally. As shown by the plots, none of the cases show significant correlation between the catalog and IceCube neutrinos. In fact, all six cases peak at $\hat{n}_s = 0$ with $\text{TS}_\text{max} = 0$.

Therefore, we set an upper limit at the 95\% confidence level for each case, relative to the background hypothesis ($n_s = 0$). According to Wilks' theorem~\cite{Wilks1938}, this corresponds to $\text{TS} \simeq -3.84$. At 100 TeV, this catalog accounts for at most 27.8\% ($\Gamma = -2.00$), 10.3\% ($\Gamma = -2.53$), and 1.7\% ($\Gamma = -3.0$) for equal weighting, and at most 12.6\% ($\Gamma = -2.0$), 4.8\% ($\Gamma = -2.53$), and 0.84\% ($\Gamma = -3.0$) for an X-band flux-density weighting.

{\it The right panel} in Fig.~\ref{fig_stacking} shows the energy fluxes versus neutrino energy ($E_\nu$). Here the red and blue lines are upper limits at the 95\% confidence level. The hardest spectrum that we choose, $\Gamma = -2.0$, is motivated by the Fermi acceleration mechanism as well as IceCube's measurements using 10 years of muon-track events between a few TeV and 10~PeV, which gives $-2.28^{+0.08}_{-0.09}$~\cite{Stettner:2019tok}. The medium spectrum, $\Gamma = -2.53$, is chosen to match IceCube's measurements using shower events from 16~TeV to 2.6~PeV~\cite{Aartsen:2020aqd} ($-2.53 \pm 0.07$) as well as earlier measurements using both track and shower events in a similar energy range ($-2.50 \pm 0.09$)~\cite{Aartsen:2015knd}. The softest spectrum, $\Gamma = -3.0$, is motivated by IceCube's measurements using high-energy starting events (HESE; $> 60$~TeV deposit energy), which gives $-2.89^{+0.20}_{-0.19}$~\cite{Abbasi:2020jmh}. 

For the energy flux ($E_\nu d F/dE_\nu$) integrated from 16 TeV to 2.6 PeV, at the 95\% confidence level, this catalog accounts for at most 30.6\% ($\Gamma = -2.00$), 10.3\% ($\Gamma = -2.53$), and 2.3\% ($\Gamma = -3.0$) for equal weighting, and at most 13.8\% ($\Gamma = -2.0$), 4.8\% ($\Gamma = -2.53$), and 1.1\% ($\Gamma = -3.0$) for X-band flux-density weighting.

Overall, the X-band flux-density weighting gives less correlation significance and neutrino flux, so the X-band flux densities of radio-bright AGN may not be an indicator of high-energy emission.

In the stacking analysis above we choose two well-motivated weighting schemes. We believe other possibly well-motivated schemes would not lead to a significant correlation, because the distribution of $\sqrt{ \text{TS}_\text{max} }$ of all the sources closely follows the standard normal distribution and none of the sources show a strong global significance (Sec.~\ref{sec_SglSrcAnlys}).

In Appendix~\ref{app_injNus}, we estimate the significance if there were signal neutrinos from these sources. We do this by injecting signal neutrinos to the current dataset (Sec.~\ref{sec_data_nus}) and performing the same stacking analysis. Our results show that even a small contribution to the total diffuse astrophysical neutrino flux from the sources would have shown significance in our analysis.

\subsection{Comparison with previous work}
\label{sec_cmpr2prevWk}

The correlation between IceCube neutrinos and the same catalog of radio-bright AGN was previously studied in Refs.~\cite{Plavin:2020emb, Plavin:2020mkf}, the first of which found a $3.1\sigma$ correlation with 56 muon-track events ($> 200$~TeV; 2009--2019).  Our analysis differs from theirs in the following ways: We include 1) the background component in a more rigorous way [Eq.~(\ref{eq_llh_func})], using the background PDF directly calculated from data,
and 2) a 2D Gaussian term [Eq.~(\ref{eq_Gaussian})], that weights the events differently in terms of their distance from the sources, instead of using the same weights within certain circular regions around the source locations. Moreover, Ref.~\cite{Plavin:2020mkf} found a $3\sigma$ correlation with a pretrial p-value sky map calculated from muon-track data from 2008--2015~\cite{IC_pvalue_web} and $4.1\sigma$ after combining the above two analyses.  We, on the other hand include the information, including direction and angular error, available for each neutrino event.

Ref.~\cite{Plavin:2020emb} found that four sources---1253-055 (3C 279), 2145+067, 1741-038, and 1730-130 (NRAO 530)---that were among the brightest in the X-band fluxes drove the significance of their signal. The p value increased from $\simeq 0.1\%$ to only $\simeq 10\%$ if the four sources were removed. Our results, however, do not show any significance from the four sources. In particular, for 3C 279, our analysis yields a local p value of only 0.7, consistent with 0.63 obtained by IceCube Collaboration also using the ten years of data~\cite{Aartsen:2019fau}. We also note that three of the four sources above, including 3C 279, were within the enlarged circles (by $0.5^\circ$) claimed to take into account IceCube's systematic uncertainties. Neither our analysis nor those done by IceCube (e.g., Refs~\cite{Abbasi:2010rd, Aartsen:2013uuv, Aartsen:2014cva, Aartsen:2015wto, Aartsen:2016oji, Aartsen:2018ywr, Aartsen:2019fau, Abbasi:2020dfi, IceCube:2018cha, IceCube:2018dnn, Abbasi:2009ig, Abbasi:2011qc, Abbasi:2012zw, Aartsen:2014aqy, Aartsen:2016qcr, Aartsen:2017wea, Aartsen:2016lir, Aartsen:2020eof}) use enlarged angular errors.

In the stacking analysis that correlates all the sources with all-sky muon-track events (Sec.~\ref{sec_StackingAnlys}), our results do not show any significance either. Moreover, our analysis shows that this catalog of radio-bright AGN could account for {\it at most} 30\% (95\% confidence level) of the total diffuse astrophysical neutrino flux, in comparison to an estimated $\sim 25\%$ contribution found in Ref.~\cite{Plavin:2020emb}.


\section{Conclusions}
\label{sec_conclusions}

High-energy astrophysical neutrinos provide a crucial window to study our Universe~\cite{Kistler:2006hp, Beacom:2007yu, Murase:2010cu, Murase:2013ffa, Murase:2013rfa, Ahlers:2014ioa, Tamborra:2014xia, Murase:2014foa, Bechtol:2015uqb, Kistler:2016ask, Bartos:2016wud, Sudoh:2018ana, Bartos:2018jco, Bustamante:2019sdb, Hovatta:2020lor} and fundamental physics~\cite{Hooper:2002yq, Borriello:2007cs, Connolly:2011vc, Klein:2013xoa, Gauld:2015kvh, Aartsen:2017kpd, Bustamante:2017xuy, Cornet:2001gy, Lipari:2001ds, AlvarezMuniz:2002ga, Beacom:2002vi, Han:2003ru, Han:2004kq, Hooper:2004xr, Hooper:2005jp, GonzalezGarcia:2005xw, Ng:2014pca, Ioka:2014kca, Bustamante:2016ciw, Salvado:2016uqu, Coloma:2017ppo, Zhou:2019vxt, Beacom:2019pzs, IceCube:2021rpz}. Fully opening the window necessitates the knowledge of their origins, for which we know little at present. Encouraging results including TXS 0506+056~\cite{IceCube:2018cha, IceCube:2018dnn} and NGC 1068~\cite{Aartsen:2019fau} have been obtained after dedicated searches by the IceCube Collaboration~\cite{Abbasi:2010rd, Aartsen:2013uuv, Aartsen:2014cva, Aartsen:2016oji, Aartsen:2019fau}. However, the majority of the diffuse astrophysical neutrinos remain unexplained, while some types of sources are being excluded as dominant neutrino emitters~\cite{Waxman:1997ti, Abbasi:2009ig, Abbasi:2011qc, Abbasi:2012zw, Aartsen:2014aqy, Murase:2016gly, Aartsen:2016qcr, Aartsen:2017wea, Aartsen:2016lir, Hooper:2018wyk, Yuan:2019ucv, Luo:2020dxa, Smith:2020oac, Senno:2017vtd, Esmaili:2018wnv, Aartsen:2020eof}. Therefore, more types of sources should be considered, including those not characterized by high-energy electromagnetic emission~\cite{Murase:2015xka, Capanema:2020rjj, Capanema:2020oet}.

In this paper, we investigate the possibility that radio-bright AGN contribute to the high-energy astrophysical neutrinos. We use an unbinned maximum-likelihood-ratio method~\cite{Braun:2008bg, Braun:2009wp, Abbasi:2010rd, Aartsen:2013uuv} and recently published IceCube muon-track data for 10 years of observation~\cite{Abbasi:2021bvk, data_webpage}.  The 3388 radio-bright AGN form a complete catalog~\cite{Plavin:2020emb} and are selected from the Radio Fundamental Catalog~\cite{RFCweb}, which collects VLBI observations since 1980 (Sec.~\ref{sec_data_srcs}). These sources show very diverse gamma-ray properties. The data comprise 1134450 muon-track events in five different construction phases of IceCube from 2008 to 2018, which have different instrumental response functions (Sec.~\ref{sec_data_nus}). The unbinned likelihood method (Sec.~\ref{sec_llh}) takes into account the information of every single event and is extensively used by IceCube collaboration~\cite{Abbasi:2010rd, Aartsen:2013uuv, Aartsen:2014cva, Aartsen:2015wto, Aartsen:2016oji, Aartsen:2018ywr, Aartsen:2019fau, Abbasi:2020dfi} and extensively used other astroparticle experiments (e.g., Refs.~\cite{Bays:2011si, Ackermann:2015lka, Ahnen:2016qkx, Abdallah:2018qtu}).

Our results are as follows. From our analysis which investigates every source location (Sec.~\ref{sec_SglSrcAnlys}), we do not find any source with large global significance. The two sources with the highest $\text{TS}_\text{max}$ have global significance of only $\simeq1.5\sigma$ and $0.8\sigma$ despite local significance of $\simeq$ $4.1\sigma$ and $3.8\sigma$ (Table~\ref{tab_list_srcs}).  Our calculated $\sqrt{ \text{TS}_\text{max} }$ of all the sources follow the standard normal distribution (Fig.~\ref{fig_TS_dist}), which implies that the background hypothesis is favored, implying that these radio-bright AGN might not be strong high-energy neutrino emitters. This is further and quantitatively supported by our stacking analysis in Sec.~\ref{sec_StackingAnlys}.

Our stacking analysis (Sec.~\ref{sec_StackingAnlys}), which seeks a correlation between the neutrinos and the sources, does not find a signal (Fig.~\ref{fig_stacking} left). All the scenarios we considered have their TS values peak at zero with zero best-fit signal event. These null results allow upper limits (95\% confidence level) to be placed on the contribution of these sources to the energy flux of the all-sky diffuse astrophysical neutrinos measured by IceCube~\cite{Aartsen:2015knd, Aartsen:2020aqd} (Fig.~\ref{fig_stacking} right). In the equal-weighting scheme that assumes high-energy neutrino emission is independent of the sources' intrinsic properties, these radio-bright AGN account for at most 30.6\% between 16 TeV and 2.6 PeV. In the X-band flux-density weighting that weights to the sources by their X-band brightness, these radio-bright AGN account for at most 13.8\%. We believe other possibly well-motivated schemes would not lead to a significant correlation, as the distribution of $\sqrt{ \text{TS}_\text{max} }$ of all the sources closely follows the standard normal distribution and none of the sources show strong global significance (Sec.~\ref{sec_SglSrcAnlys}).

Moreover, our results show that the X-band flux densities of radio-bright AGN might not be an indicator of high-energy neutrino emission because 1) the highest significant sources (some are shown in Table~\ref{tab_list_srcs}) do not have higher X-band flux densities compared to the other sources in the catalog, and 2) from the stacking analysis, the X-band flux-density weighting gives even lower upper limits on the sources' neutrino flux than the equal weighting.

The discovery prospects of high-energy neutrino sources keep increasing rapidly with more and more data to be collected by IceCube~\cite{Aartsen:2015knd}, KM3NeT~\cite{Adrian-Martinez:2016fdl}, Baikal-GVD~\cite{Avrorin:2013uyc}, P-One~\cite{Agostini:2020aar}, and especially IceCube-Gen2 (about 10 times bigger than IceCube)~\cite{Blaufuss:2015muc, Aartsen:2020fgd}. The improvements include statistics, energy range, flavor information, better reconstructions (especially angular resolution), and better sensitivities to certain parts of the sky. Meanwhile, other sources and catalogs in other wavelengths should also be tested, including those that do not show strong GeV--TeV gamma-ray emission~\cite{Murase:2015xka, Capanema:2020rjj, Capanema:2020oet}.


\section*{Acknowledgments}
We are grateful for helpful discussions with John Beacom, Mauricio Bustamante, Po-Wen Chang, Dan Hooper, Xiaoyuan Huang, Ali Kheirandish, Yuri Y. Kovalev, Ranjan Laha, Tim Linden, and especially Francis Halzen, Qinrui Liu, Kohta Murase, and Alexander Plavin. This work was supported at Johns Hopkins by NSF Grant No.\ 1818899 and the Simons Foundation. YFL was supported by the National Natural Science Foundation of China (No. U1738136).

\appendix

\section{Significance of the stacking analysis with injected signal events}
\label{app_injNus}

\begin{figure}[h!]
\includegraphics[width=\columnwidth]{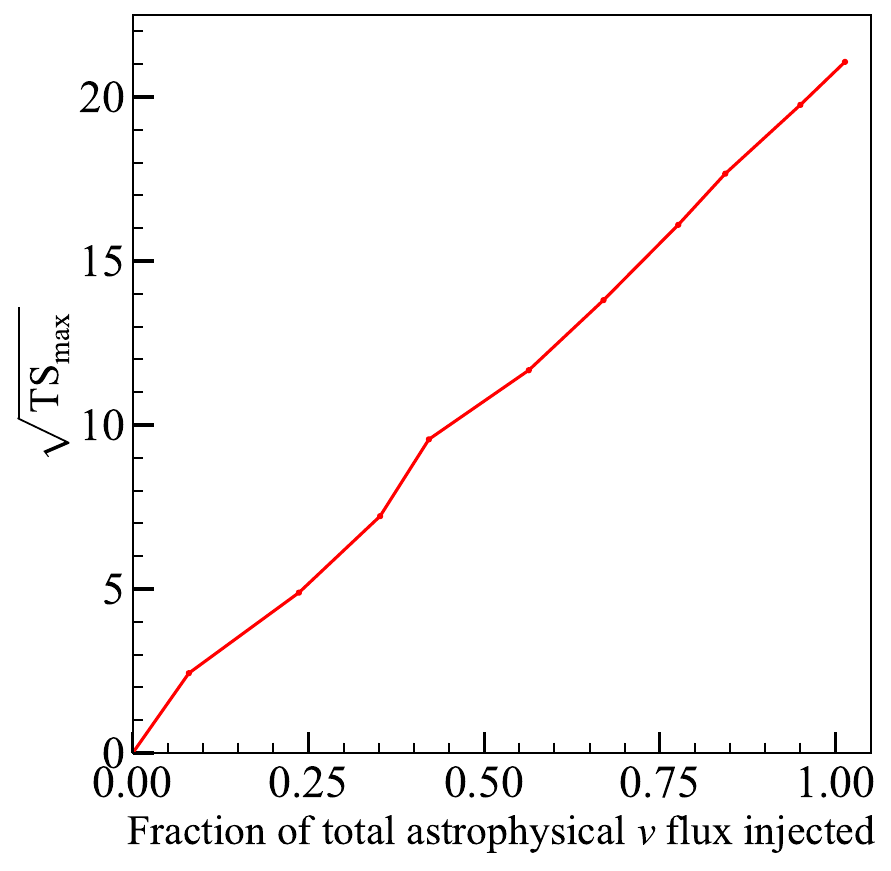}
\caption{Significance ($\simeq \sqrt{ \text{TS}_\text{max} }$) from our stacking analysis versus fraction of total astrophysical neutrino flux that we inject to these sources.  See the text for details.
}
\label{fig_TS_injectedSignals}
\end{figure}

Here we show that if the 3388 sources in the catalog contributed to the diffuse astrophysical neutrino flux observed by IceCube, our analysis would have returned a significant detection.
We do not include the simulation uncertainties. 

We start with the total diffuse astrophysical neutrino flux measured by IceCube~\cite{Aartsen:2015knd, Aartsen:2020aqd}, which is about $ d F(E_\nu) / d E_\nu =  \Phi_0 \times \left({E_\nu}/{100\ \text{TeV}}\right)^{-2.53}$, where $\Phi_0 = 4.98 \times 10^{-18}\, \rm GeV^{-1}cm^{-2}s^{-1}sr^{-1}$ is the flux normalization at 100 TeV. We then get the number of neutrinos $N_\nu$ at different energies and declinations ($= \text{zenith angles} - 90^\circ$ for IceCube location) in each phase of IceCube by  
\begin{equation}
\begin{aligned}
N^k_\nu (E_\nu, \delta) \simeq  2\pi \, t_k \, A^k_\text{eff}(E_\nu, \delta) \,  E_\nu \frac{d F}{d E_\nu}(E_\nu) ,
\label{eq_astro_nuflux_inj}
\end{aligned}
\end{equation}
where $k$ labels different phases or data samples of IceCube, just as in Sec.~\ref{sec_llh}.

Next, we assign these neutrinos to the 3388 sources with the same weight. For each neutrino with energy $E_\nu$, we use the instrumental response functions for different phases of IceCube [$\text{PDF}^k(\sigma|E_\nu)$] provided by Ref.~\cite{data_webpage} to determine its possible angular error $\sigma$. Then, the possible angular distance between the event and its assigned source is obtained by smearing the direction of the event by a 2D symmetric Gaussian distribution with $\sigma$. We mix these fake signal events with the true events in each data sample discussed in Sec.~\ref{sec_data_nus}.

Finally, we do a stacking analysis in the equal-weighting scheme just as in Sec.~\ref{sec_StackingAnlys}. We get the $\text{TS}_\text{max}$ for different fluxes of injected neutrinos, characterized by $f \times F(E_\nu)$, where $f$ is the fraction of the astrophysical neutrino flux in Eq.~(\ref{eq_astro_nuflux_inj}). The background PDFs are unchanged.

Fig.~\ref{fig_TS_injectedSignals} shows the $\sqrt{\text{TS}_\text{max}}$ ($\simeq$ significance) versus $f$. The significance increases linearly with the injected signal events because of the relation, $\text{significance} \propto N_S/\sqrt{N_B}$, where $N_S$ and $N_B$ are numbers of signal and background events, respectively. The figure shows that even a small $f$ would have provided a significant signal in our analysis. In fact, an $f = 15\%$ contribution to the total flux would have given a $\sim 3 \sigma$ signal, and a 25\% contribution would have shown $\sim 5 \sigma$. Note that $2\sigma$ corresponds to $\sim 10\%$ contribution, which is consistent with our upper limit in the case of equal weighting and $\Gamma=-2.53$ from our stacking analysis (Sec.~\ref{sec_StackingAnlys}).

\section{Results with previously released 2010--2012 data}
\label{app_3yr_data}

\begin{figure}[h!]
\includegraphics[width=\columnwidth]{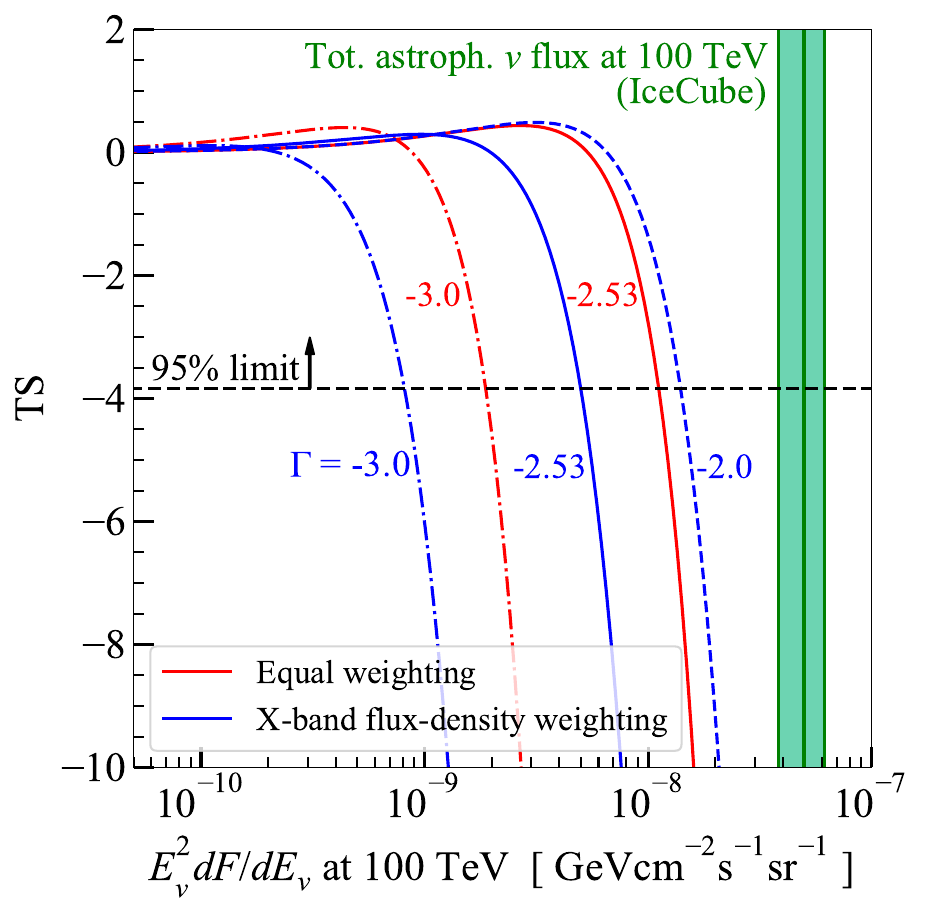}
\caption{
The same as the left panel in fig.~\ref{fig_stacking} but using previously released three years of IceCube muon-track data.
}
\label{fig_stacking_3yr} 
\end{figure}

This work initially used previously released three years of IceCube muon-track data~\cite{data3yr_webpage} (June 2010 -- May 2012) until the 10 years of data were released~\cite{Abbasi:2021bvk, data_webpage}. Here we present the results using the former dataset as a comparison.  This dataset corresponds to IC79, IC86-I, and IC86-II in the newly released dataset except for a few differences as discussed in Sec.~\ref{sec_data_nus}.

Fig.~\ref{fig_stacking_3yr} shows the result. None of the scenarios show a large significance. The highest peak has only $\text{TS}_\text{max} \simeq 0.48$. When we change this dataset to the corresponding IC79, IC86-I, and IC86-II in the new dataset, the highest peak lowers to $\text{TS}_\text{max} \simeq 0.15$ (not shown in the figure), possibly because the events originally have angular errors of less than $0.2^\circ$ were set to be $0.2^\circ$ in order to avoid unaccounted-for uncertainties and to ensure no single event dominates the likelihood analysis~\cite{data_webpage}.  The constraints on the fluxes are weaker due to fewer statistics. We do not show the case of $\Gamma = -2.0$ in the equal-weighting scheme because it overshoots the astrophysical neutrino spectrum measured by IceCube~\cite{Aartsen:2015knd, Aartsen:2020aqd} at higher energies.

\bibliography{references}

\end{document}